\documentclass[twoside]{article}

\usepackage{arxiv}

\usepackage[utf8]{inputenc} % allow utf-8 input
\usepackage[T1]{fontenc}    % use 8-bit T1 fonts
\usepackage{hyperref}       % hyperlinks
\usepackage{url}            % simple URL typesetting
\usepackage{booktabs}       % professional-quality tables
\usepackage{amsfonts}       % blackboard math symbols
\usepackage{physics}
\usepackage{nicefrac}       % compact symbols for 1/2, etc.
\usepackage{microtype}      % microtypography
\usepackage{graphicx}
\usepackage{float}
\usepackage{tikz}
\usepackage{listings}
\usepackage[backend=biber, style=phys]{biblatex}
\usepackage{doi}
\usepackage{xspace}
\usepackage{adjustbox}
\usepackage[labelfont=bf]{caption}
\usepackage{subcaption}
\usepackage{multirow}
\usepackage{wrapfig}

\usetikzlibrary{calc}
\usetikzlibrary{positioning}
\usetikzlibrary{shapes.geometric, arrows}
\usetikzlibrary{quantikz}

\tikzstyle{res} = [rectangle, rounded corners, 
minimum width=3cm, 
minimum height=3cm,
text centered, 
draw=black, 
fill=blue!30]

\tikzstyle{var} = [rectangle, 
minimum width=0cm, 
minimum height=0cm, 
text centered, 
text width=0.5cm, 
draw=white, 
fill=white!30]

\tikzstyle{arrow} = [thick,->,>=stealth]

\RequirePackage{xcolor}

\definecolor{sintefblue}{HTML}{003C65}

\definecolor{sintefcyan}{HTML}{22A7E5}
\definecolor{sintefmagenta}{HTML}{EC008C}
\definecolor{sintefgreen}{HTML}{A4C21F}
\definecolor{sintefyellow}{HTML}{F7E918}

\definecolor{sintefred}{HTML}{BE3C37}

\definecolor{sintefgrey}{HTML}{A19589}
\colorlet{sintefgray}{sintefgrey}
\definecolor{sinteflightgrey}{HTML}{D8D0C7}
\colorlet{sinteflightgray}{sinteflightgrey}

\def\oosqrt{0.7071067811865475}

\newcommand{\drawqubit}[4][green]{
    \draw[#1,thick] (#2,#3) circle (#4);
    \draw[#1,thick] (-#4+#2,#3) arc (180:360:#4 and 0.5*#4);
    \draw[#1] (-#4+#2,#3) arc (180:0:#4 and 0.5*#4);
    \draw [#1,very thick,-stealth] (#2,#3) -- (#2+\oosqrt*#4,#3+\oosqrt*#4);
}

\newcommand{\QuantumReservoirPy}{\texttt{QuantumReservoirPy}\xspace}

\title{\QuantumReservoirPy: A Software Package for Time Series Prediction}

\date{} 					% Or removing it

\author{Stanley Miao \smallskip \\
    David R. Cheriton School of Computer Science \\
    University of Waterloo \smallskip \\
    \And
    Ola Tangen Kulseng \smallskip \\
    Department of Physics \\
    Norwegian University of Science and Technology \\
	\AND
	Alexander Stasik\thanks{Corresponding author} \smallskip \\
    Department of Mathematics and Cybernetics \\
	SINTEF Digital \medskip \\
    \And
    Franz G. Fuchs \smallskip \\
    Department of Mathematics and Cybernetics \\
	SINTEF Digital \medskip \\
}

\renewcommand{\shorttitle}{QuantumReservoirPy: A Software Package for Time Series Prediction}
\renewcommand{\shortauthor}{S. Miao, O. T. Kulseng, A. Stasik, F. G. Fuchs}

\hypersetup{
pdftitle={QuantumReservoirPy: A Software Package for Time Series Prediction},
pdfsubject={quant-ph, cs.LG},
pdfauthor={Stanley~Miao, Alexander~Stasik, Franz G.~Fuchs},
}

\begin{document}
\maketitle

\begin{abstract}
	In recent times, quantum reservoir computing has emerged as a potential resource for time series prediction. Hence, there is a need for a flexible framework to test quantum circuits as nonlinear dynamical systems. We have developed a software package to allow for quantum reservoirs to fit a common structure, similar to that of reservoirpy\footnote{https://github.com/reservoirpy/reservoirpy} which is advertised as 
 ``a python tool designed to easily define, train and use [classical] reservoir computing architectures".
    Our package results in simplified development and logical methods of comparison between quantum reservoir architectures. Examples are provided to demonstrate the resulting simplicity of executing quantum reservoir computing using our software package.
\end{abstract}

\section{Introduction} \label{sec:introduction}
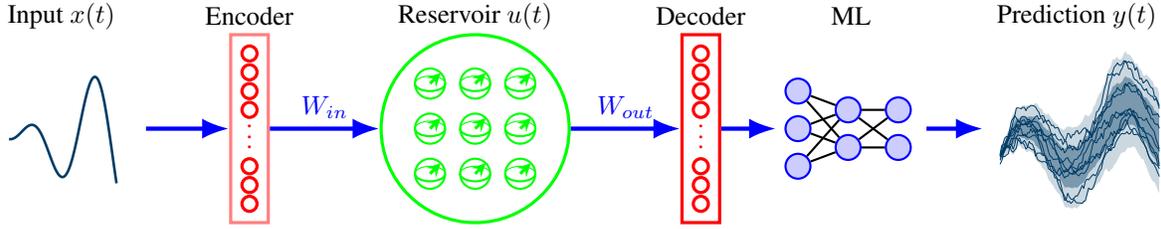
\begin{figure}
    \centering
            \begin{tikzpicture}
            % input node
            \node (rect) [draw, minimum width= .5cm, minimum height=2.5cm, inner sep=0pt, color=red!50,very thick,align=center] at (-3, 0) (IN) {};
            \node[] at (-3,1.5) {Encoder};
            
            \foreach \i in {-1,-.75,-0.5,0.25,0.5,.75,1}
            {
            \node (circle) [draw, circle, minimum size = .2cm,inner sep=0pt, color=red,very thick,align=center] at (-3, \i) {};
            }
            \node[color=red] at (-3,0) {$\vdots$};
            
            % output node
            \node (rect) [draw, minimum width= .5cm, minimum height=2.5cm, inner sep=0pt, color=red,very thick,align=center] at (3, 0) (OUT) {};
            \node[] at (3,1.5) {Decoder};
            
            \foreach \i in {-1,-.75,-0.5,0.25,0.5,.75,1}
            {
            \node (circle) [draw, circle, minimum size = .2cm,inner sep=0pt, color=red,very thick,align=center] at (3, \i) {};
            }
            \node[color=red] at (3,0) {$\vdots$};
            
            % input/output
            % \node[label=above:{Input}] at (-5.5,0) (TIN) {\includegraphics[width=2cm]{time/time_input2.jpeg}};
            % \node[label=above:{Prediction}] at (6,0)  (TOUT) {\includegraphics[width=3cm]{time/time_output2.jpeg}};
            \node[] at (-5.5,0) (TIN) {\includegraphics[width=2cm]{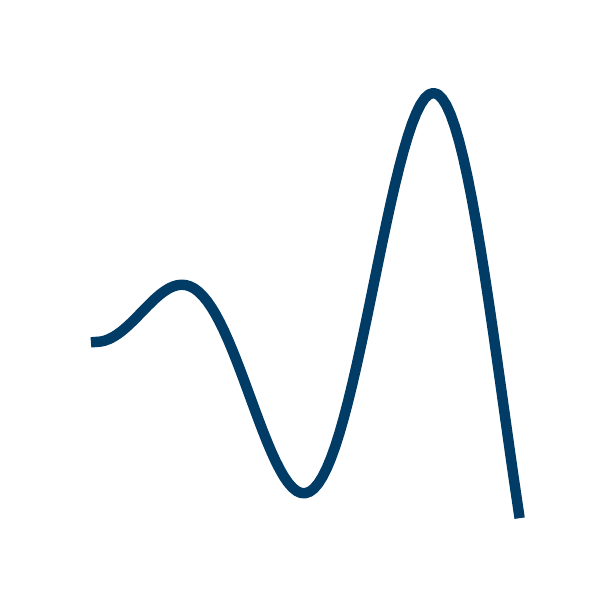}};
            \node[] at (-5.5,1.5) {Input $x(t)$};
            \node[] at (8,0)  (TOUT) {\includegraphics[width=3cm]{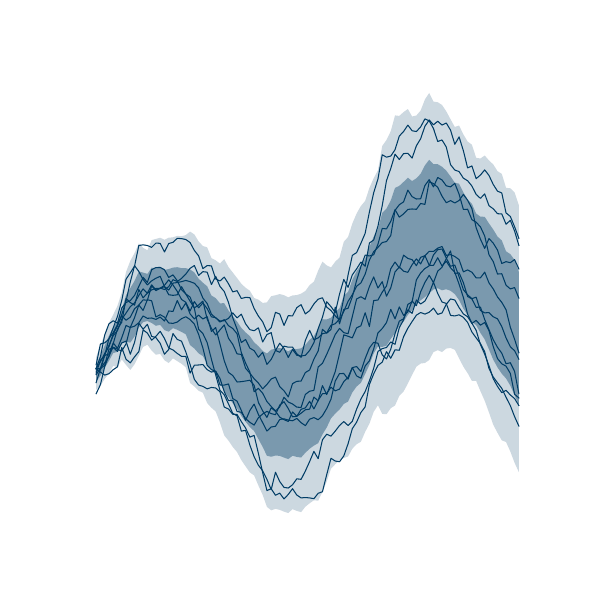}};
            \node[] at (8,1.5) {Prediction $y(t)$};
            % Reservoir
            \node (circle) [draw, circle, minimum size = 2.5cm,inner sep=-5pt, color=green,very thick,text width=2.5cm,align=center] at (0, 0) (RES) {
            };
            \node[] at (0,1.5) {Reservoir $u(t)$};
                \def\mx{-5.2}
                \def\my{-1.6}
                \drawqubit{4.6+\mx}{2.2+\my}{.2};
                \drawqubit{5.2+\mx}{2.2+\my}{.2};
                \drawqubit{5.8+\mx}{2.2+\my}{.2};
                \drawqubit{4.6+\mx}{1.6+\my}{.2};
                \drawqubit{5.2+\mx}{1.6+\my}{.2};
                \drawqubit{5.8+\mx}{1.6+\my}{.2};
                \drawqubit{4.6+\mx}{1.+\my}{.2};
                \drawqubit{5.2+\mx}{1.+\my}{.2};
                \drawqubit{5.8+\mx}{1.+\my}{.2};
                
            % Arrows    
            \draw[-Latex,ultra thick,blue] (IN) -- node [above, blue] {$W_{in}$}(RES);
            \draw[-Latex,ultra thick,blue] (RES) -- node [above, blue] {$W_{out}$}(OUT);
            % \draw[-Latex,ultra thick,green] (RES) -- node [above, green] {$W_{out}$} node [below, green] {optimized}(OUT);
            \draw[-Latex,ultra thick,blue] (TIN) -- (IN);
            \draw[-Latex,ultra thick,blue] (OUT) -- (4,0);
            \draw[-Latex,ultra thick,blue] (6,0) -- (6.75,0);
            
            \foreach \N [count=\lay,remember={\N as \Nprev (initially 0);}] in {3,2,2}{ % loop over layers
                \foreach \i [evaluate={\y=(\N/2-\i)/2+.25; \x=3.625+\lay/1.5; \prev=int(\lay-1);}] in {1,...,\N}{ % loop over nodes
                    \node[thick,draw=blue,fill=blue!20,circle] (N\lay-\i) at (\x,\y) {};
                    \ifnum\Nprev>0 % connect to previous layer
                        \foreach \j in {1,...,\Nprev}{ % loop over nodes in previous layer
                            \draw[thick] (N\prev-\j) -- (N\lay-\i);
                        }
                    \fi
                }
            }
            \node[] at (5,1.5) {ML};
            \end{tikzpicture}
    \caption{A quantum reservoir system consists of a learning task, an en- and de-coder (red) and the dynamic system itself (green). In standard RC the machine learning part is linear regression.
    }
    \label{fig:RCschematic}
\end{figure}

Reservoir computing (RC) is a paradigm in machine learning for time series prediction. With recent developments, it has shown a promising advantage in efficiency due to the relative simplicity of the associated training process over conventional neural network methods~\cite{tanaka2019recent}.
In reservoir computing, a dynamical system composed of hidden functional representations with non-linear state transitions is chosen as a reservoir. Input data from a time series is sequentially encoded and fed into the reservoir. The hidden parameters of the reservoir undergo a non-linear evolution dependent on the stored state and the encoded information fed into the system. Features are subsequently decoded from a readout of certain parameters of the reservoir, which is used to train a simple linear model for the desired time series prediction output, see Figure~\ref{fig:RCschematic} for an illustration.
The selected reservoir must incorporate non-linear state transitions within the limitations of a fixed structure. A reservoir can be virtual, such as a sparsely-connected recurrent neural network with random fixed weights, termed as an echo state network \cite{jaeger04} or even physical, such as a bucket of water \cite{fernando03}.

RC can be phrased as a supervised machine learning problem. Typically, one has an observed input sequence $\{x_t\}$ and a target sequence $\{y_t\}$. The reservoir $f$ then performs an evolution in time given by
      $
% \begin{equation}
    u_{t+1} = f(u_t, x_t)
% \label{eq:reservoir_general}
% \end{equation}
$
where $u_t$ is the internal state of the reservoir at time $t$.
% The internal state is typically very high dimensional, compared to the input and target dimension ($\dim(x), \dim(y) \ll \dim(u)$).
The output of the reservoir is given by
% \begin{equation}
$
\hat{y}_t = W_\text{out} h(u_t)
$
% \label{eq:reservoir_general_output}
% \end{equation}
where $h$ is the observation function. If the reservoir is fully observed, $h$ reduces to the identity. $W_\text{out}$ is a matrix mapping the observation on the target sequence, typically by minimizing the loss $\mathcal{L} = \sum_{t} \|(\hat{y}_t - y_t)\|_p$. For $p=2$ this reduces to linear regression with mean squared error. The performance of the reservoir is determined by the dynamical properties of the reservoir $f$.

Quantum reservoir computing (QRC) is a proposed method of reservoir computing. Multi-qubit systems with the capability of quantum entanglement provide compelling non-linear dynamics that match the requirements for a feasible reservoir. Furthermore, the exponentially-scaling Hilbert space of large multi-qubit systems support the efficiency and state-storage goals of reservoir computing. As a result, quantum computers have been touted as a viable dynamical system to produce the intended effects of reservoir computing.

In QRC, data is encoded by operating on one or more qubit(s) to reach a desired state. To obtain the desired complex non-linearity in a quantum system as a reservoir, entangling unitary operations are performed over the system. The readout is measured from one or more qubit(s) of the quantum state, which can be achieved through partial or full measurement over the system. Since quantum measurement results in a collapse to the measured state, repetitive measurements over identical systems are used to sample from the distribution of the unknown quantum state, which is then post-processed to obtain a decoded measurement for the subsequent linear model. Furthermore, this collapse results in a loss of retained information and entanglement in the system. Where only a partial measurement is taken (as in \cite{yasuda23}), this may have a desired effect of a slow leak of information driven from earlier input. When measurement is taken over the full system, the system may instead be restored through re-preparation of the pre-existing system, achieved in \cite{suzuki22} and through the restarting and rewinding protocols in \cite{mujal23}.

Existing implementations of QRC have used proprietary realizations on simulated and actual quantum computers. The lack of a shared structure between implementations has resulted in a disconnect with comparing reservoir architectures. In addition, individual implementations require a certain amount of redundant procedure prior to the involvement of specific concepts.
We observe that there is a need for a common framework for the implementation of QRC. As such, we have developed a software package, \QuantumReservoirPy, to solve the presented issues in current QRC research. In providing this software package, we hope to facilitate logical methods of comparison in QRC architecture and enable a simplified process of creating a custom reservoir from off-the-shelf libraries with minimal overhead requirements to begin development.

\section{Software Package} \label{sec:software-package}

\begin{wrapfigure}{r}{0.4\textwidth}
    \centering
    % \vspace{-5\baselineskip}
    \begin{tikzpicture}[
        resnode/.style={rectangle, very thick, minimum width=2cm, minimum height=0.75cm
        }
    ]
        \node[resnode, draw=yellow, fill=yellow!20, text=black] (before) {\texttt{before}};
        \node[resnode, draw=blue, fill=blue!10, text=black] (during) [below=0.2cm of before] {\texttt{during}};
        \node[resnode, draw=green, fill=green!10, text=black] (after) [below=0.2cm of during] {\texttt{after}};

        \node[resnode, draw=red, fill=red!10, text=black] (run) [right=.75cm of before] {\texttt{run}};
        \node[resnode, draw=orange, fill=orange!10, text=black] (predict) [right=.75cm of during] {\texttt{predict}};

        \draw[very thick] (-1.5, 2.25) rectangle (4.2, -3.);
        \draw[dashed] (-1.25,  1.1) rectangle (1.25, -2.75);
        \draw[dashed] (1.5,  1.1) rectangle (4, -2.75);

        \node at (-.25, 1.75) {\large \texttt{QReservoir}};
        \node at (-0, 0.75) {Construction};
        \node at (+2.6, 0.75) {Processing};
    \end{tikzpicture}
    \caption{
    %A structural overview of quantum reservoirs in \QuantumReservoirPy.
    Quantum circuit construction may be customized through the \texttt{before}, \texttt{during}, and \texttt{after} methods and a timeseries processed with the \texttt{run} and \texttt{predict} methods.}
    \label{fig:qreservoir-structure}
    \vspace{-2\baselineskip}
\end{wrapfigure}
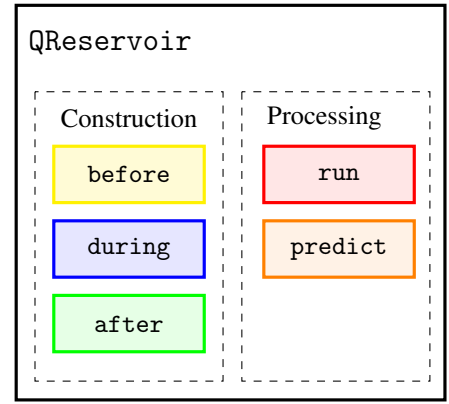
We intend \QuantumReservoirPy to provide flexibility to all possible designs of quantum reservoirs, with full control 
over pre-processing, input, quantum circuit operations, 
measurement, and post-processing. In particular, we take inspiration from the simple and flexible structure provided by the ReservoirPy software package \cite{trouvain20}.

\subsection{Structure and Design} \label{sec:structure-design}

Unlike the parameterized single-class structure of ReservoirPy, \QuantumReservoirPy uses an abstract class-based structure (see Figure~\ref{fig:qreservoir-structure}) as the former would be restrictive in the full customization of QRC. In particular, quantum circuit operations and measurement may be conducted in differing arrangements. With an abstract class, we allow the user to define the functionality of the quantum circuit, which provides full flexibility over all existing implementations of QRC. This structure also implicitly provides access to the full Qiskit circuit library\footnote{\url{https://docs.quantum.ibm.com/api/qiskit/circuit_library}}.

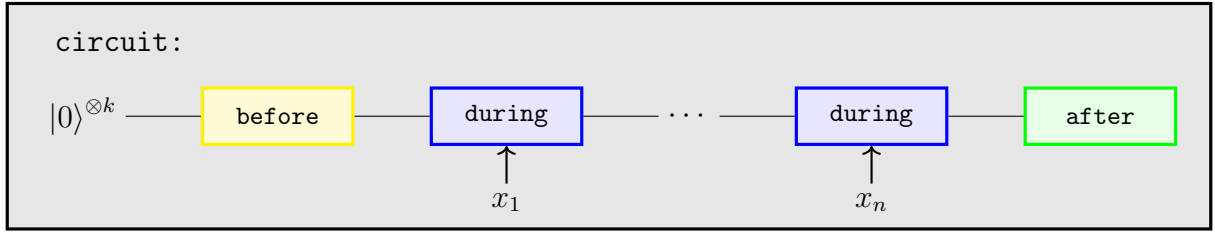
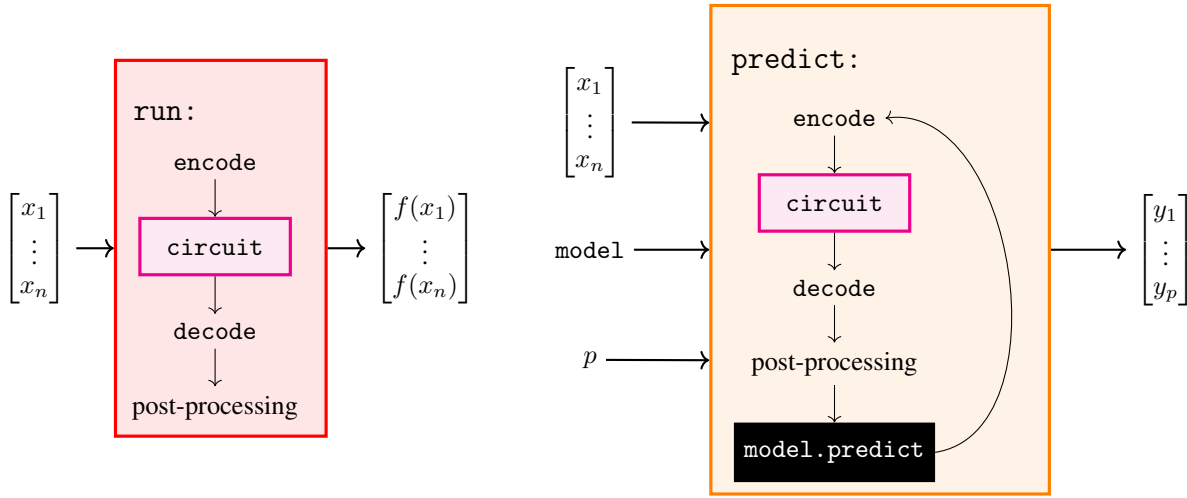
\begin{figure}
    \centering
\begin{subfigure}[b]{1\textwidth}
    \centering
    \begin{tikzpicture}[
        resnode/.style={rectangle, very thick, minimum width=2cm, minimum height=0.75cm}
    ]
        \draw[draw=black, fill=black!10, very thick] (-1, 1.5) rectangle (15, -1.5);
        \node at (0.5, 1) {\large \texttt{circuit:}};
        
        \node (init) {\large $\ket{0}^{\otimes k}$};
        \node[resnode, draw=yellow, fill=yellow!20, text=black] (before) [right=of init] {\texttt{before}};
        \node[resnode, draw=blue, fill=blue!10, text=black] (during1) [right=of before] {\texttt{during}};
        \node (ellipsis) [right=of during1] {\large $\cdots$};
        \node[resnode, draw=blue, fill=blue!10, text=black] (duringn) [right=of ellipsis] {\texttt{during}};
        \node[resnode, draw=green, fill=green!10, text=black] (after) [right=of duringn] {\texttt{after}};

        \node (x1) [below=0.5cm of during1] {\large $x_1$};
        \node (xn) [below=0.5cm of duringn] {\large $x_n$};

        \draw[-] (init.east) -- (before.west);
        \draw[-] (before.east) -- (during1.west);
        \draw[-] (during1.east) -- (ellipsis.west);
        \draw[-] (ellipsis.east) -- (duringn.west);
        \draw[-] (duringn.east) -- (after.west);

        \draw[->, thick] (x1.north) -- (during1.south);
        \draw[->, thick] (xn.north) -- (duringn.south);
    \end{tikzpicture}
    \caption{A functional overview of the hidden quantum circuit architecture common to all quantum reservoirs, where $x_t$ is the observed input sequence. The customized \texttt{before}, \texttt{during}, and \texttt{after} methods are applied sequentially to the quantum system.}
    \label{fig:circuit-function}
\end{subfigure}
    
    \begin{subfigure}[b]{.44\textwidth}
    \centering
    \begin{tikzpicture}
        \node (input) {$\begin{bmatrix} x_1 \\ \vdots \\ x_n \end{bmatrix}$};
        \node[rectangle, very thick, minimum width=2.8cm, minimum height=5cm, draw=red, fill=red!10] (run) [right=of input,xshift=-.5cm] {};
        \node (output) [right=of run,xshift=-.5cm] {$\begin{bmatrix} f(x_1) \\ \vdots \\ f(x_n) \end{bmatrix}$};

        \draw[->, thick] (input.east) -- (run.west);
        \draw[->, thick] (run.east) -- (output.west);

        \node at (2.25-.5, 1.8) {\large \texttt{run:}};
        
        \node (encode) at (2.4, 1.15) {\texttt{encode}};
        \node[rectangle, very thick, minimum width=2cm, minimum height=0.75cm, draw=magenta, fill=magenta!10] (circuit) [below=0.5cm of encode] {\texttt{circuit}};
        \node (decode) [below=0.5cm of circuit] {\texttt{decode}};
        \node (postprocess) [below=0.5cm of decode] {post-processing};

        \draw[->] (encode.south) -- (circuit.north);
        \draw[->] (circuit.south) -- (decode.north);
        \draw[->] (decode.south) -- (postprocess.north);
    \end{tikzpicture}
    \vspace{2\baselineskip}
    \caption{The input timeseries is encoded into the quantum circuit. After measurements are decoded from the quantum circuit, they are post-processed into the transformed feature vector.}
    \label{fig:run-function}
    \end{subfigure}
    \hfill
    \begin{subfigure}[b]{.54\textwidth}
    \centering
    \begin{tikzpicture}
        \node (model) {\texttt{model}};
        \node (timeseries) [above=.6cm of model] {$\begin{bmatrix} x_1 \\ \vdots \\ x_n \end{bmatrix}$};
        \node (p) [below=1cm of model] {$p$};
        \node[rectangle, very thick, minimum width=4.5cm, minimum height=6.5cm, draw=orange, fill=orange!10] (predict) [right=1.5cm of model, xshift=-.5cm] {};
        \node (output) [right=of predict] {$\begin{bmatrix} y_1 \\ \vdots \\ y_p \end{bmatrix}$};

        \draw[->, thick] (timeseries.east) -- ($(predict.west) + (0, 1.69)$);
        \draw[->, thick] (model.east) -- (predict.west);
        \draw[->, thick] (p.east) -- ($(predict.west) - (0, 1.47)$);
        \draw[->, thick] (predict.east) -- (output.west);

        \node at (2.75, 2.5) {\large \texttt{predict:}};
        
        \node (encode) at (3.25, 1.75) {\texttt{encode}};
        \node[rectangle, very thick, minimum width=2cm, minimum height=0.75cm, draw=magenta, fill=magenta!10] (circuit) [below=0.5cm of encode] {\texttt{circuit}};
        \node (decode) [below=0.5cm of circuit] {\texttt{decode}};
        \node (postprocess) [below=0.5cm of decode] {post-processing};
        \node[rectangle, very thick, minimum width=2cm, minimum height=0.75cm, draw=black, fill=black, text=white] (predict) [below=0.5cm of postprocess] {\texttt{model.predict}};

        \draw[->] (encode.south) -- (circuit.north);
        \draw[->] (circuit.south) -- (decode.north);
        \draw[->] (decode.south) -- (postprocess.north);
        \draw[->] (postprocess.south) -- (predict.north);
        \draw[->] (predict.east) to[bend right=90] (encode.east);
    \end{tikzpicture}
    \caption{The feature vector produced by encoding, decoding, and pre-processing is used by the model to predict the next step in the timeseries. This process is repeated $p$ times and returns the resulting prediction sequence.}
    \label{fig:predict-function}
    \end{subfigure}
    \caption{The intended functionality of the \texttt{run} and \texttt{predict} method.
    The observed input sequence is $\{x_t\}$ and the target sequence $\{y_t\}$. The reservoir $f$ performs an evolution in time.
    }
\end{figure}

The construction methods in \QuantumReservoirPy serve as the sequential operations performed on the quantum system. The operations in the \texttt{before} method prepares the qubits, which may include initialization to a default, initial, or previously saved state. The \texttt{during} method provides the operations that are subsequently applied for each step in the timeseries. This may include (but is not limited to) measurement, re-initialization, and entangling operations. Finally, the operations in the \texttt{predict} method are applied once following the processing of the entire timeseries, which may include the transformation of final qubit states and measurement. Figure~\ref{fig:circuit-function} demonstrates the aforementioned arrangement, which is implemented as a hidden intermediary process in a \QuantumReservoirPy quantum reservoir.

The processing methods serve as functions acting on the quantum reservoir itself. The \texttt{run} method is used to process training data by taking a timeseries as input and returning the transformed data after being processed by the quantum reservoir. This is done using the hidden \texttt{circuit} interface as presented in Figure~\ref{fig:circuit-function}, where data encoding and decoding follow the implementation of the custom construction methods. Depending on the realization of QRC, such as averaging over multi-shot data, additional post-processing is included in the \texttt{run} method to achieve the desired output. The transformed data from the \texttt{run} method serves as training data for a simple machine learning model. Figure~\ref{fig:run-function} provides a visualization of the \texttt{run} method.

The \texttt{predict} method functions as a complete forecasting process involving the same hidden \texttt{circuit} interface, encoding, decoding, and post-processing. Additionally, a trained simple machine learning model is used to predict the next step in the timeseries from the transformed and post-processed data. The resulting prediction is then fed in as input for the following prediction, which occurs as an iterative process until the specified number of forecasting steps is reached. At this point, the \texttt{predict} method returns the sequence of predictions from each iteration. Figure~\ref{fig:predict-function} provides a visualization of the \texttt{predict} method.

\subsection{Dependencies} \label{sec:dependencies}

The three main dependencies of \QuantumReservoirPy are numpy, qiskit, and scikit-learn.
We strive for \QuantumReservoirPy to support compatibility with existing reservoir computing and quantum computing workflows.
\QuantumReservoirPy uses NumPy as a core dependency. Allowed versions of NumPy ensure compatibility with classical ReservoirPy to facilitate comparison between classical and quantum reservoir architectures\footnote{An example is provided in the public GitHub repository under the \texttt{examples/static/} subdirectory.}.
Much of existing research in QRC is performed on IBM devices and simulators (see \cite{yasuda23}, \cite{suzuki22}), programmed through the Qiskit software package. To minimize disruption in current workflows, \QuantumReservoirPy is built as a package to interact with Qiskit circuits and backends. It is expected that the user also use Qiskit in the customization of reservoir architecture when working with \QuantumReservoirPy.
The model used by the \texttt{predict} function is expected to be a scikit-learn estimator. Likewise to our dependency on Qiskit, this allows for complete customization over the choice of model and faster adoption of \QuantumReservoirPy from a simple package structure.

\subsection{Distribution} \label{sec:distribution}
The software package is developed and maintained through a public GitHub repository\footnote{\url{https://github.com/OpenQuantumComputing/quantumreservoirpy/}} provided through the \texttt{OpenQuantumComputing} organization. The \texttt{main} branch of this repository serves as the latest stable version of \QuantumReservoirPy. Installation through cloning this repository is intended for development purposes.
Official releases of the software package are published through the Python Package Index (PyPI) under the \texttt{quantumreservoirpy} name. Installation through PyPI is the suggested method of installation for general-purpose use.
\QuantumReservoirPy is licensed under the GNU General Public License v3.0. \QuantumReservoirPy also includes derivative work of Qiskit, which is licensed by IBM under the Apache License, Version 2.0.

\subsection{Documentation} \label{sec:documentation}

Documentation is mainly provided online in the form of a user guide and API reference. The user guide includes steps for installation and getting started with a simple quantum reservoir. Examples are also provided for further guidance. The API reference outlines the intended classes and functions exposed by the software package.
A brief overview of the software package is also included as a \texttt{README} file at the top level of the public GitHub repository.

\section{Usage} \label{sec:usage}

\QuantumReservoirPy provides two prepackaged reservoirs that implement the processing methods \texttt{run} and \texttt{predict} according to the structure presented in section \ref{sec:structure-design}. These reservoirs use common QRC schemes as a basis to get started with the software package or restrict customization to circuit construction. As such, the construction of a reservoir using either of these schemes only requires user implementation of the sequential circuit methods \texttt{before}, \texttt{during}, and \texttt{after}.

\subsection{\texttt{Static} Reservoirs} \label{sec:static-reservoirs}

The \texttt{Static} abstract reservoir class is structured to run multi-shot processing according to a repeated-measurement process. A single circuit is created according to the circuit construction methods and the timeseries parameter. Measurement is expected in the \texttt{during} function to generate the transformed reservoir output using the single circuit. This circuit is sent to the specified Qiskit backend and is run with the remaining parameters provided to the processing methods. The resulting measurement data is post-processed by taking the average over all shots before it is returned to the user as decoded reservoir output.

An example of a \texttt{Static} quantum reservoir is the following.
\begin{lstlisting}[language=Python]
class QuantumReservoir(Static):

def before(self, circuit):
    circuit.h(circuit.qubits)

def during(self, circuit, timestep):
    circuit.initialize(encoder[timestep], [0, 1])
    circuit.append(operator, circuit.qubits)
    circuit.measure([0, 1])

\end{lstlisting}
In the example, the timestep passed to the \texttt{during} method is encoded in the circuit according to an \texttt{encoder}. Measurement is also taken in the \texttt{during} method to provide sequential data using a single circuit for the entire timeseries. The \texttt{after} method is not necessary for this scheme and is left as the inherited empty method by default.

\subsection{\texttt{Incremental} Reservoirs} \label{sec:incremental-reservoirs}

The \texttt{Incremental} abstract reservoir class processes data using multiple circuits over a moving substring of the timeseries. Circuits of length at most the specified \texttt{memory} parameter are created for step-by-step processing of the timeseries according to the circuit construction methods. Each circuit with a fixed-length memory is sent to the Qiskit backend to be run with the remaining parameters to the processing methods. Using this scheme, measurements taken in the \texttt{after} method can be post-processed to provide the user with the desired reservoir output.

An example of an \texttt{Incremental} quantum reservoir is the following.
\begin{lstlisting}[language=Python]
class QuantumReservoir(Incremental):

def before(self, circuit):
    circuit.h(circuit.qubits)

def during(self, circuit, timestep):
    circuit.initialize(encoder[timestep], [0, 1])
    circuit.append(operator, circuit.qubits)

def after(self, circuit):
    circuit.measure_all()

\end{lstlisting}
In the example, the \texttt{during} method is still used to encode the current timestep in the timeseries, but measurement is instead taken in the \texttt{after} method. Since a circuit is created for each timestep, the combined measurements produce the desired sequential data.

\subsection{Custom Reservoirs} \label{sec:custom-reservoirs}

Custom reservoirs provide full flexibility over processing data in a quantum reservoir. A custom reservoir can be created by implementing the \texttt{QReservoir} abstract class. Unlike the prepackaged reservoirs, a custom reservoir must implement the \texttt{run} and \texttt{predict} methods, in addition to the circuit construction methods.

\subsection{Processing} \label{sec:quantum-backends}

When a quantum reservoir is instantiated, it requires a Qiskit backend. If no backend is specified, then \texttt{AerSimulator} is used by default. The backend must support the circuit operations specified in the circuit construction methods. When creating a custom reservoir, the \texttt{run} and \texttt{predict} methods should run the quantum reservoir on \texttt{self.backend} attribute.

As an example, instantiation of a quantum reservoir using the simulator backend \texttt{FakeTorontoV2} is as follows.
\begin{lstlisting}[language=Python]
from qiskit.providers.fake_provider import FakeTorontoV2

backend = FakeTorontoV2()
reservoir = QuantumReservoir(n_qubits=4, backend=backend)
\end{lstlisting}

Once a quantum reservoir has been instantiated, training data for the trainable model is produced by using the \texttt{run} method on the timeseries. The trained model can then be passed to the \texttt{predict} method to make predictions.

\begin{lstlisting}[language=Python]
output = reservoir.run(timeseries=timeseries, shots=10000)
# ...(model training)...
predictions = reservoir.predict(num_pred=10, model=model, from_series=timeseries, shots=10000)
\end{lstlisting}
Processing of a timeseries using a quantum reservoir to predict 10 timesteps. Model training of a scikit-learn estimator between the \texttt{run} and \texttt{predict} methods is not shown. The \texttt{shots} parameter is directly passed to the Qiskit backend when using a prepackaged reservoir.

\subsection{Example}
\begin{figure}
    \centering
    \begin{minipage}{1\textwidth}
    \centering
    \begin{subfigure}[b]{.49\textwidth}
        \centering
        \begin{lstlisting}[language=Python, linewidth=1\linewidth, basicstyle=\small\ttfamily]
class QuantumReservoir(Static):
    def during(self, circuit, timestep):
        circuit.initialize(timestep, 0)
        circuit.append(operator, circuit.qubits)
        circuit.measure(0)
       \end{lstlisting}% \vspace{5px} \\
       \caption{Implementation of a basic quantum reservoir.}
       \label{sfiga}
    \end{subfigure}
    \hfill
    \begin{subfigure}[b]{.49\textwidth}
        \centering
        \includegraphics[width=1\linewidth]{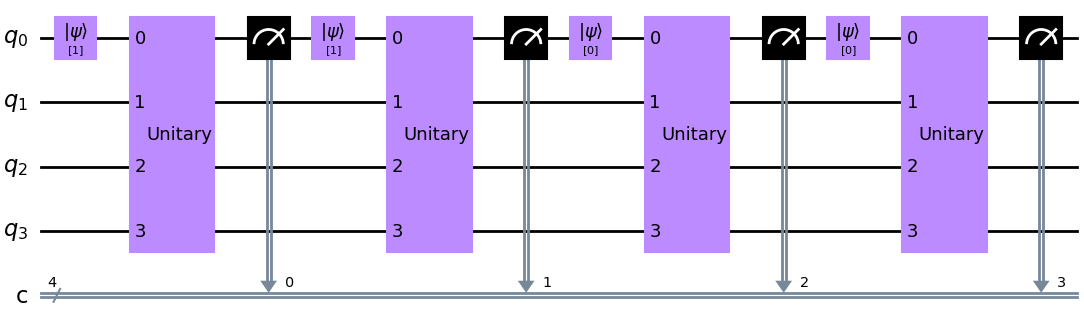}
        \caption{Resulting circuit produced by the implementation above.}
       \label{sfigb}
    \end{subfigure}
    \end{minipage}

    \begin{minipage}{.75\textwidth}
    \centering
    \begin{subfigure}[b]{1\textwidth}
        \centering
        \includegraphics[width=1.0\linewidth]{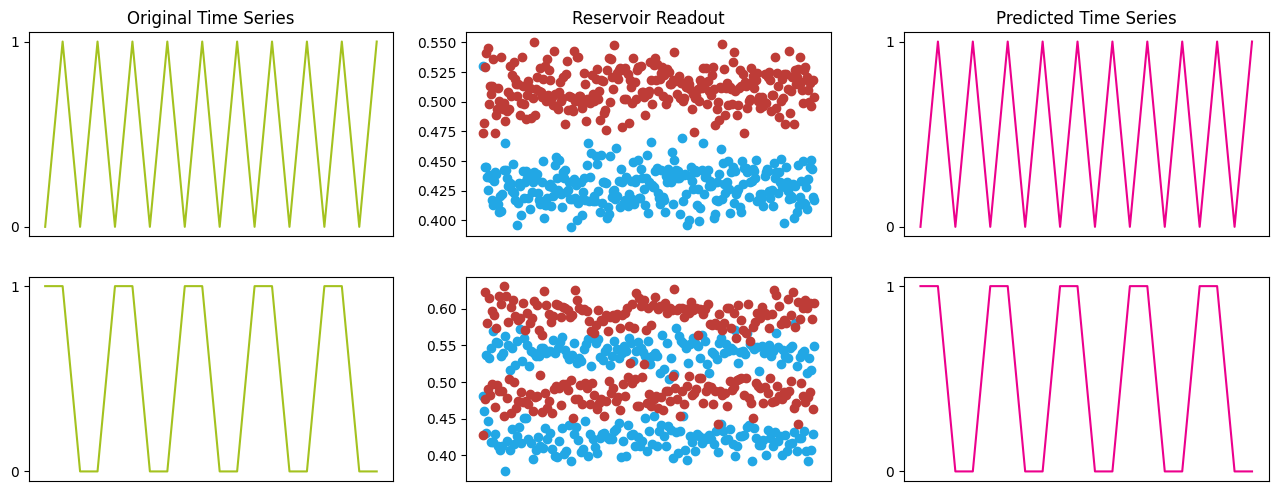}
        \caption{Plotted training sequences, reservoir feature spaces, and resulting predictions of some quantum reservoirs. %The results in the top two rows were produced by the quantum reservoir on the left.}
        }
       \label{sfigc}
    \end{subfigure}
    \end{minipage}
    \caption{A simple example of the usage of \QuantumReservoirPy for time series prediction.}
    \label{fig:simple-overview}
\end{figure}

An example for how a quantum reservoir can be used for predictions is given in Figure~\ref{fig:simple-overview}.
A \texttt{Static} quantum reservoir with four qubits is used to predict a discrete sequence of zeros and ones.
The backend is an ideal \texttt{AerSimulator}.
One can see that the reservoir readout can be differentiated with linear regression, which leads to very good predictions.

\section{Conclusion} \label{sec:conclusion}

We have introduced a new software package to encapsulate the possibilities of quantum reservoir computing within a succinct and easy-to-use framework. The structure has been designed to provide the full suite of the Qiskit circuit library when choosing a reservoir layout involving timeseries encoding, reservoir dynamics, and qubit measurement. Post-processing on decoded measurements and prediction generation are provided as built-in and customizable features for the implementation of a quantum reservoir. Prepackaged quantum reservoirs with common processing schemes are included.

\subsection{Further Development} \label{sec:further-development}

The authors continue to support and maintain the project. Users may report package issues and desired features by opening an issue on the public GitHub repository or contacting the authors by email.
Additional opportunities for further development on \QuantumReservoirPy include supplementary built-in processing schemes, expanded features for data visualization, and reservoir evaluation methods.

\section*{Acknowledgements}

Work in this project was supported by the NTNU and SINTEF Digital through the International Work-Integrated-Learning in Artificial Intelligence (IWIL AI) program, in partnership with SFI NorwAI and the Waterloo Artificial Intelligence Institute (Waterloo.AI).
IWIL AI is funded by the Norwegian Directorate for Higher Education and Skills (HK-dir).

\printbibliography

\end{document}